\documentclass[utf8]{frontiersSCNS} 

\usepackage{url,hyperref,lineno,microtype,subcaption}
\usepackage[onehalfspacing]{setspace}
\usepackage{pgfplots}
\usepackage{tikzscale}
\usetikzlibrary{patterns}
\usepackage{booktabs}
\usepackage{xspace}
\usepackage{paralist}
\usepackage{wrapfig}
\usepackage{optidef}
\usepackage{glossaries}
\usepackage[english]{babel}

\makeatletter
\let\ORIbbl@fixname\bbl@fixname
\def\bbl@fixname#1{%
  \@ifundefined{languagealias@\expandafter\string#1}
    {\ORIbbl@fixname#1}
    {\edef\languagename{\@nameuse{languagealias@#1}}}%
}
\newcommand{\definelanguagealias}[2]{%
  \@namedef{languagealias@#1}{#2}%
}
\newcommand{\added}[1]{{#1}}
\makeatother

\definelanguagealias{en}{english}

\newcommand\arcsec{\mbox{$^{\prime\prime}$}}%

\newlength\figureheight
\newlength\figurewidth

\def\keyFont{\fontsize{8}{11}\helveticabold }
\def\firstAuthorLast{Douglas {et~al.}} 
\def\Authors{Ewan S. Douglas\,$^{1,*}$, Kevin Tracy\,$^{2}$ and Zachary Manchester\,$^{2}$}

\def\Address{$^{1}$ Department of Astronomy/Steward Observatory, 933 North Cherry Avenue, Rm. N204, Tucson, AZ 85721-0065\\
$^{2}$The Robotics Institute, Carnegie Mellon University, 5000 Forbes Ave, Pittsburgh, PA 15213}

\def\corrAuthor{Corresponding Author}

\def\corrEmail{douglase@arizona.edu}

\newacronym{AU}{AU}{astronomical Unit [1.5e11 m]}  
\newacronym{pc}{pc}{parsec}
\newacronym{mas}{mas}{milliarcsecond}
\newacronym{nm}{nm}{nanometer}
\newacronym{CTE}{CTE}{coefficient of thermal expansion}
\newacronym{sqarc}{$as^2$}{square arcsecond}

\newacronym{smc}{SMC}{Small Magellanic Cloud}
\newacronym{lmc}{LMC}{Large Magellanic Cloud}
\newacronym{ism}{ISM}{interstellar medium}
\newacronym{mw}{MW}{Milky Way}
\newacronym{epseri}{$\epsilon$ Eri}{Epsilon Eridani}
\newacronym{EKB}{EKB}{Edgeworth-Kuiper Belt}

\newacronym{CFR}{CFR}{Complete Frequency Redistribution}

\newacronym{nasa}{NASA}{National Aeronautics and Space Agency}
\newacronym{esa}{ESA}{European Space Agency}
\newacronym{omi}{OMI}{\textit{Optical Mechanics Inc.}}
\newacronym{gsfc}{GSFC}{\gls{nasa} Goddard Space Flight Center}
\newacronym{stsci}{STScI}{Space Telescope Science Institute}
\newacronym{nsroc}{NSROC}{\gls{nasa} Sounding Rocket Operations Contract}
\newacronym{wff}{WFF}{\gls{nasa} Wallops Flight Facility}
\newacronym{wsmr}{WSMR}{White Sands Missile Range}

\newacronym{irac}{IRAC}{Infrared Array Camera}
\newacronym[plural=CCDs, firstplural=charge-coupled devices (CCDs)]{ccd}{CCD}{charge-coupled device}
\newacronym[plural=EMCCDs, firstplural=electron multiplying charge-coupled devices (EMCCDs)]{EMCCD}{EMCCD}{electron multiplying charge-coupled device}

\newacronym{DM}{DM}{Deformable Mirror}
\newacronym{MCP}{MCP}{ Microchannel Plate }
\newacronym{ipc}{IPC}{Image Proportional Counter}
\newacronym{cots}{COTS}{Commercial Off-The-Shelf}
\newacronym{ISR}{ISR}{incoherent scatter radar}
\newacronym{atcamera}{AT}{angle tracker}
\newacronym{MEMS}{MEMS}{microelectromechanical systems}
\newacronym{QE}{QE}{quantum efficiency}
\newacronym{RTD}{RTD}{Resistance Temperature Detector}
\newacronym{PID}{PID}{Proportional-Integral-Derivative}
\newacronym{PRNU}{PRNU}{photo response non-uniformity}
\newacronym{DSNU}{PRNU}{dark signal non-uniformity}
\newacronym{CMOS}{CMOS}{complementary metal–oxide–semiconductor}
\newacronym{TRL}{TRL}{technology readiness level}
\newacronym{swap}{SWaP}{Size, Weight, and Power}
\newacronym{ConOps}{ConOps}{concept of operations}
\newacronym{NRE}{NRE}{non-recurring engineering}
\newacronym{CBE}{CBE}{current best estimate}

\newacronym{FOV}{FOV}{field-of-view}
\newacronym{NIR}{NIR}{near-infrared}
\newacronym{PV}{PV}{Peak-to-Valley}
\newacronym{MRF}{MRF}{Magnetorheological finishing}
\newacronym{AO}{AO}{Adaptive Optics}
\newacronym{TTP}{TTP}{tip, tilt, and piston}
\newacronym{FPS}{FPS}{fine pointing system}
\newacronym{SHWFS}{SHWFS}{Shack-Hartmann Wavefront Sensor}
\newacronym{OAP}{OAP}{off-axis parabola}
\newacronym{LGS}{LGS}{laser guide star}
\newacronym{WFCS}{WFCS}{wavefront control system}
\newacronym{OPD}{OPD}{optical path difference}
\newacronym{UA}{UA}{University of Arizona}
\newacronym{MEL}{MEL}{Master Equipment List}
\newacronym{LEO}{LEO}{low-earth orbit}
\newacronym{GEO}{GEO}{geosynchronous orbit}
\newacronym{EFC}{EFC}{electric-field conjugation}
\newacronym{LDFC}{LDFC}{linear dark field control}
\newacronym{DAC}{DAC}{digital-to-analog converter}
\newacronym{FEA}{FEA}{finite element analysis}

\newacronym{SiC}{SiC}{Silicon Carbide}
\newacronym{ESPA}{ESPA}{EELV Secondary Payload Adapter}
\newacronym{EEID}{EEID}{Earth-equivalent insolation distance}
\newacronym{LLOWFS}{LLOWFS}{Lyot Low-order wavefront sensor}
\newacronym{WFSC}{WFSC}{wavefront sensing and control}
\newacronym{STOP}{STOP}{Structural-Thermal-Optical-Performance}

\newacronym{resel}{resel}{resolution element}

\newacronym{acs}{ACS}{Attitude Control System}
\newacronym{orsa}{ORSA}{Ogive Recovery System Assembly}
\newacronym{gse}{GSE}{Ground Station Equipment}
\newacronym{FSM}{FSM}{Fast Steering Mirror}

\newacronym{WFS}{WFS}{wavefront sensor}
\newacronym{LSI}{LSI}{Lateral Shearing Interferometer}
\newacronym{VVC}{VVC}{Vector Vortex Coronagraph}
\newacronym{VNC}{VNC}{Visible Nulling Coronagraph}
\newacronym{CGI}{CGI}{Coronagraph Instrument}
\newacronym{IWA}{IWA}{Inner Working Angle}
\newacronym{OWA}{OWA}{Outer Working Angle}
\newacronym{NPZT}{N-PZT}{Nuller piezoelectric transducer}
\newacronym{ZWFS}{ZWFS}{Zernike wavefront sensor}
\newacronym{SPC}{SPC}{Shaped Pupil Coronagraph}
\newacronym{HLC}{HLC}{Hybrid-Lyot Coronagraph}
\newacronym{ADI}{ADI}{angular differential imaging}
\newacronym{RDI}{RDI}{reference differential imaging}

\newacronym{HST}{HST}{Hubble Space Telescope}
\newacronym{GPS}{GPS}{Global Positioning System}
\newacronym{ISS}{ISS}{International Space Station}
\newacronym[description=Advanced CCD Imaging Spectrometer]{acis}{ACIS}{Advanced \gls{ccd} Imaging Spectrometer}
\newacronym{stis}{STIS}{\textit{Space Telescope Imaging Spectrograph}}
\newacronym{mcp}{MCP}{Microchannel Plate}
\newacronym{jwst}{JWST}{$\textit{James Webb Space Telescope}$}
\newacronym{fuse}{FUSE}{$\textit{FUSE}$}
\newacronym{galex}{GALEX}{$\textit{Galaxy Evolution Explorer}$}
\newacronym{spitzer}{Spitzer}{$\textit{Spitzer Space Telescope}$}
\newacronym{mips}{MIPS}{Multiband Imaging Photometer for \gls{spitzer}}
\newacronym{gissmo}{GISSMO}{Gas Ionization Solar Spectral Monitor}
\newacronym{iue}{IUE}{International Ultraviolet Explorer}
\newacronym{spinr}{SPINR}{$\textit{Spectrograph for Photometric Imaging with Numeric Reconstruction}$}
\newacronym{imager}{IMAGER}{$\textit{Interstellar Medium Absorption Gradient Experiment Rocket}$}
\newacronym{TPF-C}{TPF-C}{Terrestrial Planet Finder Coronagraph}
\newacronym{RAIDS}{RAIDS}{Atmospheric and Ionospheric Detection System }
\newacronym{mama}{MAMA}{Multi-Anode Microchannel Array}
\newacronym{ATLAST}{ATLAST}{Advanced Technology Large Aperture Space Telescope}
\newacronym{PICTURE}{PICTURE}{Planet Imaging Concept Testbed Using a Rocket Experiment}
\newacronym{LITES}{LITES}{Limb-imaging Ionospheric and Thermospheric
Extreme-ultraviolet Spectrograph}
\newacronym{LBT}{LBT}{Large Binocular Telescope}
\newacronym{LBTI}{LBTI}{Large Binocular Telescope Interferometer}
\newacronym{KIN}{KIN}{Keck Interferometer Nuller}
\newacronym{SHARPI}{SHARPI}{Solar High-Angular Resolution Photometric Imager}
\newacronym{IRAS}{IRAS}{Infrared Astronomical Satellite}
\newacronym{HARPS}{HARPS}{High Accuracy Radial velocity Planetary}
\newacronym{hstSTIS}{STIS}{Space Telescope Imaging Spectrograph}
\newacronym{spitzerIRAC}{IRAC}{Infrared Array Camera}
\newacronym{spitzerMIPS}{MIPS}{Multiband Imaging Photometer for Spitzer}
\newacronym{spitzerIRS}{IRS}{Infrared Spectrograph}
\newacronym{CHARA}{CHARA}{Center for High Angular Resolution Astronomy}
\newacronym{wfirst-afta}{WFIRST-AFTA}{Wide-Field InfraRed Survey
Telescope-Astrophysics Focused Telescope Assets}
\newacronym{GPI}{GPI}{Gemini Planet Imager}
\newacronym{WFIRST}{WFIRST}{Wide-Field InfraRed Survey Telescope}
\newacronym{HabEx}{HabEx}{Habitable Exoplanet Observatory Mission Concept}
\newacronym{LUVOIR}{LUVOIR}{Large UV/Optical/Infrared Surveyor}
\newacronym{FGS}{FGS}{Fine Guidance Sensor}
\newacronym{STIS}{STIS}{Space Telescope Imaging Spectrograph}
\newacronym{MGHPCC}{MGHPCC}{Massachusetts Green High Performance
Computing Center}
\newacronym{WISE}{WISE}{Wide-field Infrared Survey Explorer}
\newacronym{ALMA}{ALMA}{Atacama Large Millimeter Array}
\newacronym{GRAIL}{GRAIL}{Gravity Recovery and Interior Laboratory}
\newacronym{jwstNIRCam}{NIRCam}{near-\gls{IR}-camera}
\newacronym{jwstMIRI}{MIRI}{Mid-Infrared Instrument}

\newacronym{AURIC}{AURIC}{The Atmospheric Ultraviolet Radiance Integrated Code} 
\newacronym{FFT}{FFT}{Fast Fourier Transform  }
\newacronym{MODTRAN}{MODTRAN   }{ MODerate resolution atmospheric TRANsmission }
\newacronym{idl}{IDL}{$\textit {Interactive Data Language}$}
\newacronym[sort=NED,description=NASA/IPAC Extragalactic Database]{ned}{NED}{\gls{nasa}/\gls{ipac} Extragalactic Database}
\newacronym{iraf}{IRAF}{Image Reduction and Analysis Facility}
\newacronym{wcs}{WCS}{World Coordinate System}
\newacronym{pegase}{PEGASE}{$\textit{Projet d'Etude des GAlaxies par Synthese Evolutive}$}
\newacronym{dirty}{DIRTY}{$\textit{DustI Radiative Transfer, Yeah!}$}
\newacronym{CUDA}{CUDA}{Compute Unified Device Architecture}
\newacronym{KLIP}{KLIP}{Karhunen-Lo\`{e}ve Image Processing}

\newacronym{MSIS}{MSIS}{Mass Spectrometer Incoherent Scatter Radar}
\newacronym{nmf2}{$N_m$}{F2-Region Peak density}
\newacronym{hmf2}{$h_m$}{F2-Region Peak height}
\newacronym{H}{$H$}{F2-Region Scale Height}

\newacronym{isr}{ISR}{Incoherent Scatter Radar}
\newacronym[description=TLA Within Another Acronym]{twaa}{TWAA}{\gls{tla} Within Another Acronym}
\newacronym[plural=SNe, firstplural=Supernovae (SNe)]{sn}{SN}{Supernova}
\newacronym{EUV}{EUV}{Extreme-Ultraviolet }
\newacronym{EUVS}{EUVS}{\gls{EUV} Spectrograph}
\newacronym{F2}{F2}{Ionospheric Chapman F Layer }
\newacronym{F10.7}{F10.7}{ 10.7 cm radio flux [10$^{-22}$ W m$^{-2}$ Hz$^{-1}$]  }
\newacronym{FUV}{FUV}{far-ultraviolet }
\newacronym{IR}{IR}{infrared}
\newacronym{MUV}{MUV}{mid-ultraviolet }
\newacronym{NUV}{NUV}{near-ultraviolet }
\newacronym{O$^+$}{O$^+$}{Singly Ionized Oxygen  Atom }
\newacronym{OI}{OI}{Neutral Atomic Oxygen Spectroscopic State }
\newacronym{OII}{OII}{Singly Ionized Atomic Oxygen Spectroscopic State }
\newacronym{PSF}{PSF}{point spread function}
\newacronym{$R_E$}{$R_E$}{Earth radii [$\approx$ 6400 km]  }
\newacronym{RV}{RV}{radial velocity}
\newacronym{UV}{UV}{ultraviolet }
\newacronym{WFE}{WFE}{wavefront error}
\newacronym{sed}{SED}{spectral energy distribution}
\newacronym{nir}{NIR}{near-infrared}
\newacronym{mir}{MIR}{mid-infrared}
\newacronym{ir}{IR}{infrared}
\newacronym{uv}{UV}{ultraviolet}
\newacronym[plural=PAHs, firstplural=Polycyclic Aromatic Hydrocarbons (PAHs)]{pah}{PAH}{Polycyclic Aromatic Hydrocarbon}
\newacronym{obsid}{OBSID}{Observation Identification}
\newacronym{SZA}{SZA}{Solar Zenith Angle}
\newacronym{TLE}{TLE}{Two Line Element set}
\newacronym{DOF}{DOF}{degrees-of-freedom}
\newacronym{PZT}{PZT}{lead zirconate titanate}
\newacronym{ADCS}{ADCS}{attitude determination and control system}
\newacronym{COTS}{COTS}{commercial off-the-shelf}
\newacronym{CDH}{C$\&$DH}{command and data handling}
\newacronym{EPS}{EPS}{electrical power system}

\newacronym{PCA}{PCA}{principal component analysis}
\newacronym{fwhm}{FWHM}{full-width-half maximum}
\newacronym{RMS}{RMS}{root mean squared}
\newacronym{RMSE}{RMSE}{root mean squared error}
\newacronym{MCMC}{MCMC}{Marcov chain Monte Carlo}
\newacronym{DIT}{DIT}{discrete inverse theory}
\newacronym{SNR}{SNR}{signal-to-noise ratio}
\newacronym{PSD}{PSD}{power spectral density}
\newacronym{NMF}{NMF}{non-negative matrix factorization}

\begin{document}
\onecolumn
\firstpage{1}
\title[Limits on Nanosatellite Pointing]{Practical limits on Nanosatellite Telescope Pointing: The Impact of Disturbances and Photon Noise} 

\author[\firstAuthorLast ]{\Authors} 
\address{} 
\correspondance{} 

\extraAuth{}
\maketitle

\begin{abstract}
Accurate and stable spacecraft pointing is a requirement of many astronomical observations.
Pointing particularly challenges nanosatellites because of an unfavorable surface area to mass ratio and proportionally large volume required for even the smallest attitude control systems.
This work explores the limitations on astrophysical attitude knowledge and control in a regime unrestricted by actuator precision or actuator-induced disturbances such as jitter. 
The external disturbances on an archetypal 6U CubeSat are modeled and the limiting sensing knowledge is calculated from the available stellar flux and grasp of a telescope within the available volume.
These inputs are integrated using a model-predictive control scheme.
For a simple test case at 1 Hz, with an 85 mm telescope and a single 11th magnitude star, the achievable body pointing is predicted to be {0.39} arcseconds.
For a more general limit, integrating available star light, the achievable attitude sensing is approximately 1 milliarcsecond, which leads to a predicted body pointing accuracy of 20 milliarcseconds after application of the control model.
These results show significant room for attitude sensing and control systems to improve before astrophysical and environmental limits are reached.

\tiny
 \keyFont{ \section{Keywords:} attitude sensing and control, environmental disturbances, CubeSats, astrophysics} 
\end{abstract}

\section{Introduction}\label{sec:intro}

{Astronomical observation with nanosatellites requires a level of precision that far exceeds what is capable from common \glspl{ADCS} present on Earth-imaging CubeSats.} Note {that} this work will refer to nanosatellites in an inclusive sense for any spacecraft approximately 10 kg or below, or {those} that are designed specifically to a CubeSat deployer standard  (\url{https://www.cubesat.org}).
Since {astronomical light} sources are dim and relatively static,  {these} fine-pointing nanosatellites {need} to observe continuously, or ``stare'', for minutes to hours (e.g. \cite{weiss_brite-constellation:_2014,shkolnik_verge_2018,Knapp20}). 
An ideal astronomical \gls{ADCS} would {achieve this} by slewing the {nanosatellite} to the target attitude, and maintaining perfect inertial pointing indefinitely {while rejecting} transient disturbances {that fall} below the sampling accuracy of the instrument. 
Improving nanosatellite pointing enables a range of applications beyond the transformative photometry recently demonstrated by the ASTERIA mission \cite{knapp_demonstrating_2020}. These {applications} span from direct exoplanet detection by interferometry \citep{dandumont_exoplanet_2020} or starshades \citep{macintosh_miniature_2019}, to X-rays \citep{krizmanic_vtxo_2020}, and the ultraviolet \citep{shkolnik_verge_2018}. 

{Actuating} the {attitude} of a spacecraft requires application of a torque {on the body of the spacecraft}. 
In early nanosatellites this was achieved with {onboard} magnetorquers {interacting} with the Earth's magnetic field, allowing {for} coarse aiming of solar panels and antenna. 
Miniaturization {and implementation} of reaction wheels,  flywheels spun continuously using precision motors at tens to thousands of revolutions per minute, {on these nanosatellites} has enabled sub-1 arcminute nanosatellite pointing \citep{sinclair_enabling_2007,mason_minxss_2016}.

\begin{figure}
    \setlength{\figureheight}{2.0in}
    \setlength{\figurewidth}{5.8in}
%
%
\definecolor{mycolor1}{rgb}{0.00000,0.44700,0.74100}%
\begin{tikzpicture}

\begin{axis}[%
width=0.951\figurewidth,
height=\figureheight,
at={(0\figurewidth,0\figureheight)},
scale only axis,
xmode=log,
xmin=0.4,
xmax=1000000,
xminorticks=true,
xlabel style={font=\color{white!15!black}},
xlabel={Spacecraft Mass (kg)},
ymin=0,
ymax=1,
ylabel style={font=\color{white!15!black}},
ylabel={ACS Mass Fraction},
axis background/.style={fill=white}
]
\addplot [color=mycolor1, line width=2.0pt, forget plot]
  table[row sep=crcr]{%
0.446683592150963	1.00581886160683\\
0.601993548138357	0.845537502887892\\
0.811304105116383	0.71090387451905\\
1.09339103884784	0.597813905525556\\
1.47355838124505	0.502820245754969\\
1.98590826684075	0.423027179429793\\
2.67639999507463	0.356002354310408\\
3.60697271532629	0.299702635686024\\
4.86110155173032	0.252411824982096\\
6.55128556859551	0.212688344444731\\
8.82913926906353	0.179321293161103\\
11.898992864271	0.15129353486039\\
16.0362212973666	0.127750692291361\\
21.6119461900249	0.107975103024346\\
29.1263265490874	0.091363942765723\\
39.2534245081376	0.0774108493123757\\
52.9016706936678	0.0656904869844781\\
71.2953531379435	0.0558455810102942\\
96.0844395349596	0.0475760266288721\\
129.492584220526	0.0406297409206466\\
174.516596540161	0.0347949784999899\\
235.195263507096	0.0298938768270206\\
316.971640937555	0.0257770343787758\\
427.181311649225	0.0223189564047676\\
575.710408926781	0.0194142294387359\\
775.882431904704	0.016974307953272\\
1045.65340282899	0.0149248152040925\\
1409.22257533746	0.0132032759849152\\
1899.20317905333	0.0117572121799313\\
2559.54792269953	0.0105425430601166\\
3449.49168201215	0.00952224155921015\\
4648.86504321484	0.00866520556828069\\
6265.25534261275	0.00794530984226644\\
8443.65756872834	0.00734060961751679\\
11379.4808414324	0.00683267166400677\\
15336.0771877001	0.00640601238052006\\
20668.3649970003	0.00604762580411353\\
27854.6662501044	0.00574658714605655\\
37539.6134148643	0.00549371976873059\\
50591.9748843582	0.00528131545186174\\
68182.5860701605	0.00510289942089082\\
91889.3768001957	0.00495303297478148\\
123838.916289267	0.00482714769671976\\
166897.172684574	0.00472140619390788\\
224926.598881409	0.00463258512134821\\
303132.605966755	0.00455797692380173\\
408530.503982962	0.0044953073006963\\
550574.795978489	0.0044426658780466\\
742007.274882455	0.00439844797404034\\
1000000	0.00436130568311636\\
};
\addplot [color=red, only marks, mark size=4.0pt, mark=*, mark options={solid, fill=red, red}, forget plot]
  table[row sep=crcr]{%
419725	0.00262076359521115\\
3300	0.0121212121212121\\
120	0.0416666666666667\\
3	0.333333333333333\\
};
\node[above right, align=left]
at (axis cs:3,0.333) {  3U CubeSat};
\node[above right, align=left]
at (axis cs:120,0.042) {  LEO SmallSat};
\node[above right, align=left]
at (axis cs:3300,0.012) {  GEO CommSat};
\node[above right, align=left]
at (axis cs:419725,0.003) {  ISS};
\end{axis}

\begin{axis}[%
width=8.167in,
height=5.194in,
at={(0in,0in)},
scale only axis,
xmin=0,
xmax=1,
ymin=0,
ymax=1,
axis line style={draw=none},
ticks=none,
axis x line*=bottom,
axis y line*=left
]
\end{axis}
\end{tikzpicture}%
    \caption{Approximate mass fraction consumed by momentum actuators as a function of spacecraft mass.} 
    \label{fig:scaling}
    \end{figure}
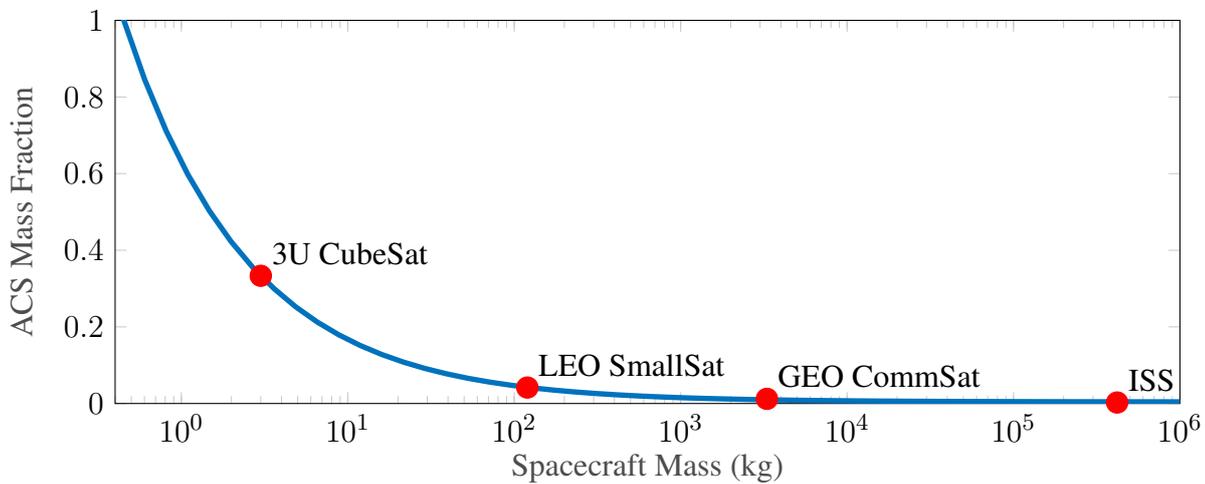

The attitude control performance of {nanosatellites} is limited by an unfavorable  mass-to-surface-area ratio of {both} the spacecraft and the small reaction wheels, increasing sensitivity to external disturbances and internal imbalances. 
While these actuators have proven capable of fine pointing on larger spacecraft, they do not scale down well to smaller spacecraft, as {demonstrated} in Fig. \ref{fig:scaling}.
State-of-the-art reaction wheels allow {for precise} pointing and rejection of slowly changing disturbances, e.g. drag differentials, but {imperfections in the reaction wheels} can directly add uncontrolled jitter and excite spacecraft structural modes   \citep{shields_characterization_2017,addari_experimental_2017}. 
Thermal drifts and misalignment between attitude sensor star-trackers and science payloads {can be addressed} when using the science telescope for {attitude determination}, but the higher-order jitter remains, e.g. analysis in \cite{smith_exoplanetsat:_2010,nguyen_fine-pointing_2018}. 

Large spacecraft often use control-moment gyros for {attitude control}, but their increased \gls{swap} and complexity make them impractical for most nanosatellites  \citep{votel_comparison_2012}. {Spacecraft like the} largest NASA observatories {are able to achieve} the most exquisite pointing via passive damping {of these actuators}.
This {damping} comes at the cost of significant mass and volume, and becomes more difficult as spacecraft mass decreases. 
Using great observatories as examples, HST uses viscous dampers \citep{davis_hubble_1986}, each Chandra reaction wheel is isolated by six damping springs arranged in a hexapodal configuration \citep{pendergast_use_1998}, and JWST uses dual stage passive isolation \citep{bronowicki_vibration_2006}.  
This simple, but fundamental trade between pointing accuracy and \gls{swap} is a major barrier to astrophysics with nanosatellites. 

Many proposed solutions to the attitude control limitations of CubeSats have been in the form of second-stage correction \citep{beierle_two-stage_2018,pong_-orbit_2018,cahoy_cubesat_2019}.
ASTERIA (Arcsecond Space Telescope Enabling Research in Astrophysics) is the first sub-arcsecond imaging CubeSat \citep{pong_-orbit_2018}. 
The correction was accomplished by image-plane stabilization using a detector-shifting \gls{PZT} stage. {The science telescope on}
ASTERIA operates at 20 Hz, reading out 50 ms exposures for science and pointing control, {limiting} the detectable stellar magnitude to $m_{\rm V}<$7 since dimmer stars do not flip the detector's first analog-to-digital bit \citep{Knapp20} in an exposure.
Similarly, the CLICK free-space laser communications CubeSats \citep{cahoy_cubesat_2019}, due for launch in 2021, will use  \gls{MEMS}-based steering mirrors to achieve fine pointing while the DeMi mission launched in 2020 is designed to use a MEMS deformable mirror for fine wavefront steering \citep{morgan_mems_2019}.

The combination of controller bandwidth and photon noise from stars presents a fundamental limit to attitude control with reaction wheels.
Without a large aperture and corresponding large number of photons per exposure, photoelectron shot noise limits the centroid accuracy of the star tracker or astronomical telescope. 
As a Poisson process, error is proportional to the square root of the number of photoelectrons per exposure; thus, for a simple disturbance environment the achievable pointing decreases linearly with telescope diameter.
Unfortunately, performance degrades faster {than this in practice}. 
Crossing a {structural mode of the spacecraft} can have devastating impacts on stability, and manufacturing tolerance limits mean reaction wheel imbalances are an even larger fraction of the spacecraft inertia {on} nanosatellites, so higher bandwidths may be required when fewer photons are available.
This work seeks to define the sensing and control limits of satellite pointing, establishing the lower limit \textit{without} considering the limitations imposed by {actuators}. 
Actuator-disturbances {are} usually so domin{ant} that environmental disturbances are entirely neglected in spacecraft design (e.g. \cite{choueiri_cost-effective_2018}).

The two-stage actuator approach is limited for several reasons: pointing correction stages add mass, power, complexity, and risk{,} not all astronomical optical layouts can accommodate a correction stage{,} and the bandwidth of the system must be sufficient to correct high-frequency reaction-wheel disturbances. 
Single-stage approaches that minimize high-frequency noise have also been proposed, including electrospray thrusters \citep{mier-hicks_electrospray_2017}, viscously damped reaction wheels \citep{underwood_using_2015}, {and} predictive magnetorquers \citep{gatherer_magnetorquer-only_2018}. 
While both approaches have various limitations in their actuation precision and dynamic range, they hold great promise. 
However, we find it imperative to assess the possible gains of these approaches by finding the fundamental limits placed by the input disturbances and sensing available on a nanosatellite platform.
This provides a context and methodology for evaluating the limits on new controller technologies.
 Sec. \ref{sec:methods} introduces our disturbance, sensing, and control models. {Sec. \ref{sec:estimation} details the state-estimation framework used in the closed-loop simulations.}  
Sec. \ref{sec:results} presents the resulting spacecraft body point{ing} as a function of sensor and control limitations. 
Sec. \ref{sec:conclusions} discusses the impact of these results, comparing them to the limitations faced by current actuators and the goals of future space astronomy missions.
\begin{table}[]
    \centering
    \begin{tabular}{c|c|c}
        Symbol & Variable & Notes/values \\
        \hline
         $D_x$ & Sensing Aperture Diameter & variable \\
         $R_N$ & Detector Read Noise & 2.6 e$^-$, \citep{micron_mt9p031_2006} \\ 
         $D_N$ & Detector Dark Noise & 25 e$^-$/s, \citep{micron_mt9p031_2006} \\ 
         $\lambda$ & Sensing wavelength & 550 nm\\ 
         $BW$ & Sensing bandwidth & 100 nm\\ 
$\tau$ & effective quantum efficiency and throughput & 0.25\\
$N_{pix}$ & Number of pixels for centroiding & 4 (minimum)\\
$f_0$ & Zero-mag flux & Johnson V\\
$m_l$ & limiting magnitude & 13\\
    \end{tabular}
    \caption{Nomenclature and assumed values.}
    \label{tab:nomenclature}
\end{table}

\section{Methods}\label{sec:methods}

\subsection{Disturbances}

Modern attitude control systems rely on flywheel-based momentum actuators like reaction wheels and control-moment gyroscopes. 
For a constant density, spacecraft inertia scales like the fifth power of spacecraft length, while disturbances like solar pressure, drag, and friction scale with {the second power of spacecraft length (surface area)}.
As {shown in figure \ref{fig:scaling}}, the mass and volume fraction consumed by momentum actuators increases {dramatically} as {the size of the} spacecraft shrinks.

This work neglects magnetic torques, which can be large \citep{inamori_jitter_2013}, but depend on the fine details of spacecraft design, since both the intrinsic magnetic moment and the dynamic magnetic moment due to on-board currents can be minimized with careful electrical design and spacecraft magnetic cleanliness (e.g. \cite{mehlem_multiple_1978,stern_techniques_2004,junge_prediction_2011,belyayev_practical_2016,lassakeur_magnetic_2020}). 

\begin{figure}[h!]
    \centering
    \setlength{\figureheight}{2.5in}
    \setlength{\figurewidth}{5.0in}
    \input{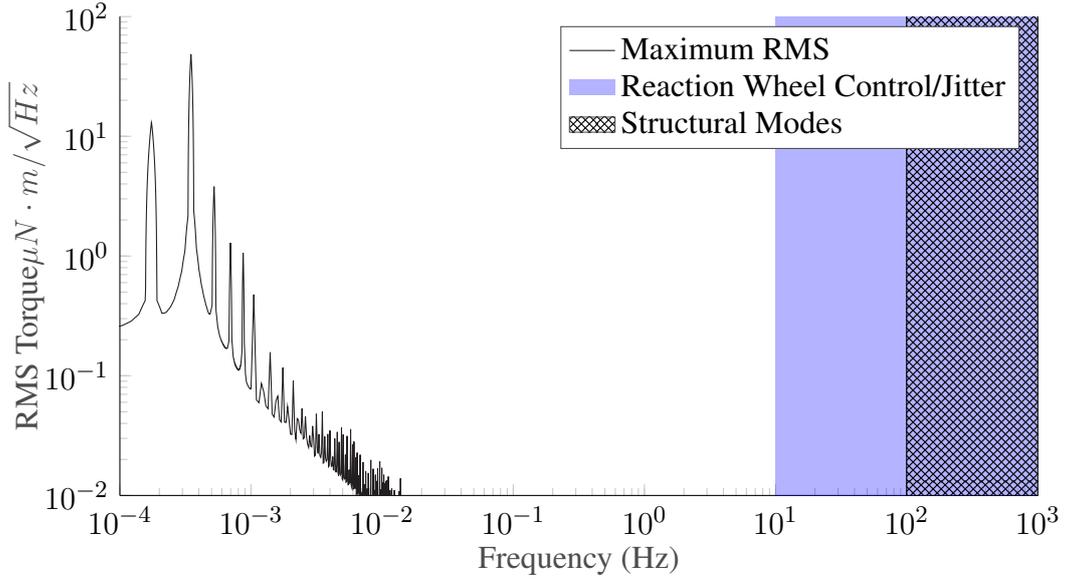}
    \caption{Frequency content of perturbation torques on a small satellite in low-Earth orbit. Disturbances are due to atmospheric drag, solar radiation pressure, and {a} gravity gradient. 
     The disturbances manifest on the left as solid curve representing the maximum value from an ensemble of orbits drawn from {multiple years} of credible low-Earth orbits. 
    On the far right, solid shading illustrates typical actuator-induced error frequencies and hatching represents typical {nanosatellite} structural modes.}
    \label{fig:disturbances}
\end{figure}

Figure \ref{fig:disturbances} depicts the frequency content of environmental disturbances acting on a 6U spacecraft in low-Earth orbit. This figure was generated by simulating 1000 Monte-Carlo runs that included drag, solar radiation pressure, and gravity gradient torques \citep{Markley14,Wertz78}. The orbits were sampled uniformly from the range of all possible lower-earth orbits with eccentricity less than 0.03, and altitudes between 400 km and 600 km.  The simulation epochs were sampled uniformly between 2014 and 2017 to avoid any bias in solar activity or 3rd body perturbations. The line in the figure is the result of the \textit{maximum} RMS torque over all Monte-Carlo runs for each given frequency. Note that the environmental disturbances that the control system must counteract are concentrated in the first few harmonics of the orbital frequency{,} and lie almost entirely in the range 0.1 to 10 milliHertz.

\begin{figure}
    \centering
    \includegraphics[width=.95\columnwidth]{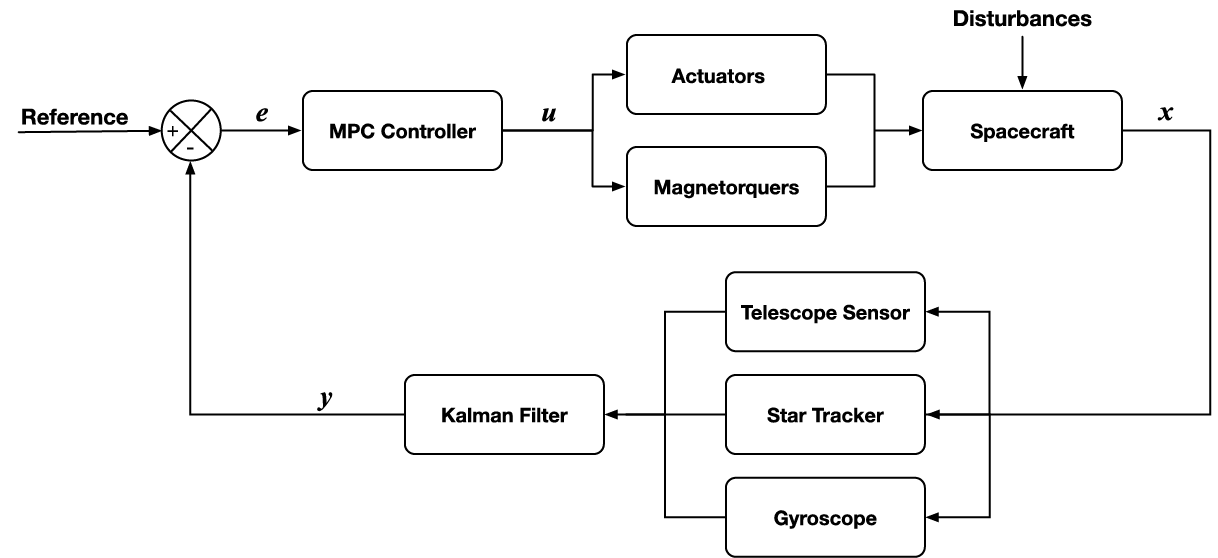}
    \caption{Control system block diagram.}
    \label{fig:control_block_diagram}
\end{figure}

In addition to poor scaling for small spacecraft, reaction wheels also produce unwanted high-frequency jitter that is often the main source of pointing error on nanosatellites. This jitter is caused by small mass imbalances in the flywheels as they rotate at hundreds of Hertz. Notably, the frequency content of this jitter is several orders of magnitude higher than the environmental disturbances the wheels are supposed to counteract.
The overall control loop for the system discussed here is depicted in Fig. \ref{fig:control_block_diagram}.
Two notable features which will be detailed subsequently, are the model-predictive controller and the addition of a telescope sensor to increase sensor precision.

\subsection{Control Methodology}

To effectively reason about actuator trade-offs and constraints, the control problem will be formulated as a constrained optimization problem and solved online in a model-predictive control (MPC) scheme. 
 Recent advances in algorithm development and microprocessors have enabled state-of-the-art high-performance MPC solvers that can be run on embedded systems such as drones and CubeSats \citep{Howell19,Jackson21,Tracy20}. 
In typical MPC implementations, the problem in Eq. \eqref{eq:mpc} is solved with a horizon of 10-100 time steps at rates between 10 Hz and 1 kHz. While MPC has been in use in industrial applications since the 1980s, its use has historically been limited by the ability to solve the necessary optimization problems at real-time rates on available computing hardware. Thanks to Moore's law, it has become possible to do this on ever more complicated systems, and today MPC is deployed in a wide range of aerospace and robotics applications, including SpaceX's autonomous rocket landings \citep{Blackmore16}, Boston Dynamics' humanoid robots \citep{Kuindersma16}, and autonomous cars \citep{Beal2013}.

Model-predictive control problems take the general form,
\begin{mini}[2]
	{x_{1:N},u_{1:N-1}}{\ell_N(x_N) + \sum_{k=1}^{N-1} \ell_k(x_k,u_k) }{}{}
	\addConstraint{x_{k+1}}{ = f(x_k,u_k)}{}
	\addConstraint{g_k(x_k,u_k)}{\leq 0}{}
	\label{eq:mpc},
\end{mini}
where $x_k$ and $u_k$ are the state and control inputs of the system at time step $k$, $\ell(x,u)$ is a cost function that penalizes deviations from a desired {reference}, $f(x,u)$ is a {discrete-time} dynamics model, and $g(x,u)$ is a set of constraints on the system, including actuator limits and safety constraints. 
In our case, $f(x,u)$ will encode the attitude dynamics of the spacecraft, the actuators, and the magnetic torque coils, $\ell(x,u)$ will penalize deviations from the desired pointing target, as well as excessive control effort, and $g(x,u)$ {will enforce torque limits} associated with candidate actuator systems.

\subsection{Sensing} 
Attitude sensing using astronomical sources requires sufficient photon counts to  determine direction by accurately measuring a star, or stars', position(s) on a sensor. 
Here we shall assume an astronomical telescope is included in the  nanosatellite payload, either as the primary science instrument, or as an adjuvant co-aligned with another sensor. 
Various means exist of pulling out attitude knowledge from a telescope and employing a separate science camera, e.g. beamsplitters, dichroics, and field-slicing. 
We will neglect the details of the implementation except to limit our sensing to a single filter, $BW=100$ nm,  leaving most incoming light available for another specialized science sensor.

In the ideal case, the angular attitude knowledge $\Delta \phi$ perpendicular to an axis ($x$) depends on the wavelength ($\lambda$), and the dimension(s) of the photon collecting aperture, ($\Delta_x$).
As shown by \cite{lindegren_astrometric_2005}, the fundamental pointing error depends on the uncertainty in the momentum of each incident photon. 
By Heisenberg's uncertainty principle ($\Delta x\Delta p\geq \hbar/2$), the photon's momentum uncertainty ($\Delta p$) decreases as uncertainty in the location of the photon increases.  
Thus, a larger telescope increases $\Delta x$ and better constrains the angle of the incident photon.

\begin{figure}
    \centering
    \includegraphics{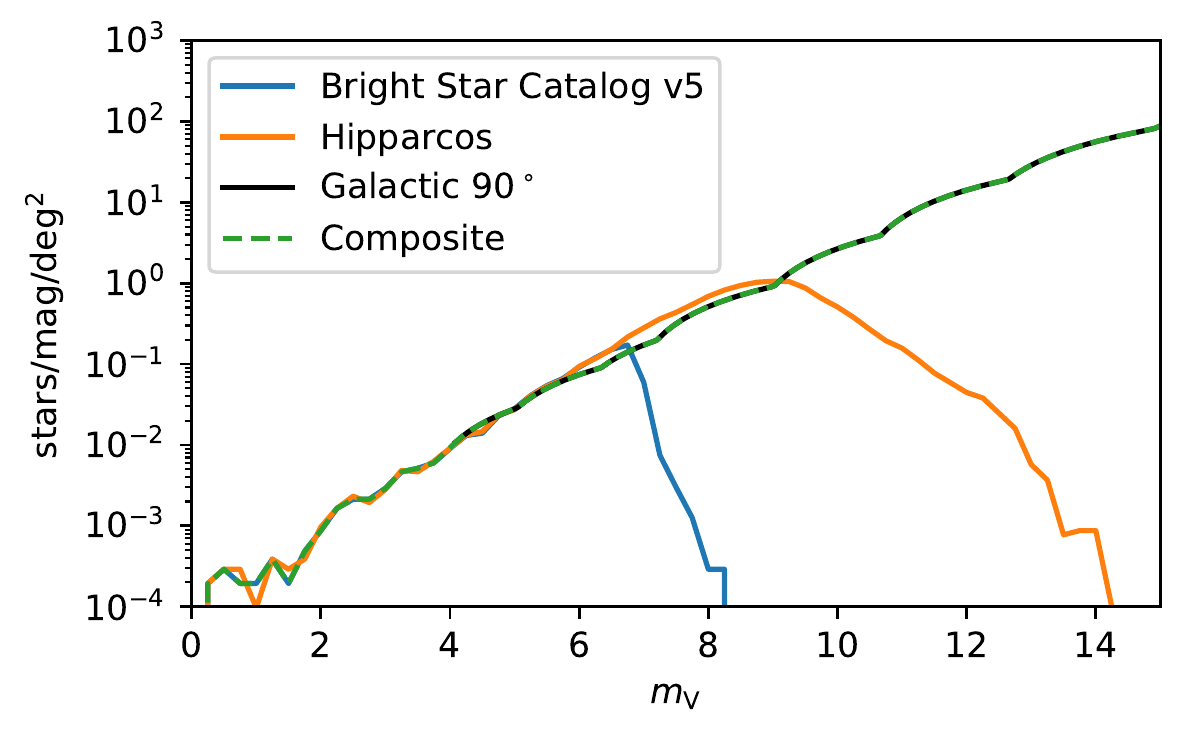}
    \caption{Star counts in units of stellar density per magnitude and per square degree. The dashed {curve} represents the density used in this work and represents a composite of the Bright Star Catalog \citep{vizier:V/50} for bright stars and the  galactic pole curve from \cite{bahcall_universe_1980} at dimmer magnitudes provides a lower limit on stellar density by assuming observations are of the sparsest region of the sky.
    This composite represents a conservative estimate of the number photons available to localize spacecraft attitude control.
    }
    \label{fig:star_density}
\end{figure}

For the case where there are $N$ photons, \cite{lindegren_high-accuracy_2013}, Eq. 16.1 gives:

\begin{equation}
   \Delta\phi = \frac{{1}}{4\pi}\frac{\lambda}{\Delta x\sqrt{N}}.
\end{equation}
Where $N=\sigma^2$ is the variance in the photons from Poisson statistics.  For a circular aperture diameter $D$, $\Delta x=D/4$ then
\begin{equation}
   \Delta\phi =  \frac{{1}}{\pi}\frac{\lambda}{D_x\sqrt{N}}.
\end{equation}
The same relation can be derived by assuming a Poisson process and Fraunhofer diffraction \cite[eq. 32]{lindegren_photoelectric_1978}.
The number of photons, $N$, received by the sensor depends on the exposure time and the light grasp or \textit{\'{e}tendue} of the system. 
The \textit{\'{e}tendue} is defined as the product of the collecting area, $\pi (Dx/2)^2$ and the solid-angle, $\Omega$ subtended by the instrument. 
A complicating factor is that {the} number of stars visible within the collecting area varies with direction {on} the sky. 
To calculate the distribution of stars on the sky, we conservatively estimate the differential number of stars per square degree $\mathcal{A}(stars/mag/deg^2$){,} where the stellar density is lowest at the galactic poles from \cite[fig 4a]{bahcall_universe_1980}.
The assumed stellar density versus magnitude is shown in Fig. \ref{fig:star_density} and is a composite of the Bright Star catalog and the galactic pole estimate.
At lower galactic latitudes, stars are more plentiful and more flux would be available than is assumed here, though deleterious effects such as confusion and reddening become more pronounced. 
The Hipparcos catalog \cite{esa_hipparcos_1997} is shown as a solid orange line for comparison, the integrated sky appearing slightly brighter until the catalog completeness falls off above  $ m_{\rm V}\sim9$. 
Newer catalogs such as Gaia \citep{mora_gaia:_2016} would provide increased precision but not discernibly alter the shape of the stellar density function for these relative bright stars.
Neglecting detector effects, integration over the distribution of stellar magnitudes ($m$) and the instrument \gls{FOV} gives the total number of photons per second received by the detector from stars up to a limiting magnitude $m_l$.
Since only four stars in the terrestrial sky are brighter than apparent magnitude zero, we neglect negative magnitudes and the total flux received is
\begin{equation}
\mathcal{F}=\int_{FOV} \int_0^{m_l}  \mathcal{A} \tau f_0 10^{-2m/5} d\Omega dm {,}
\end{equation}
where $f_0$ is the zero-point magnitude of the instrument band-pass and $\tau$ is the effective throughput including losses due to optics and sensor quantum efficiency.

\begin{figure}
    \centering
    \includegraphics{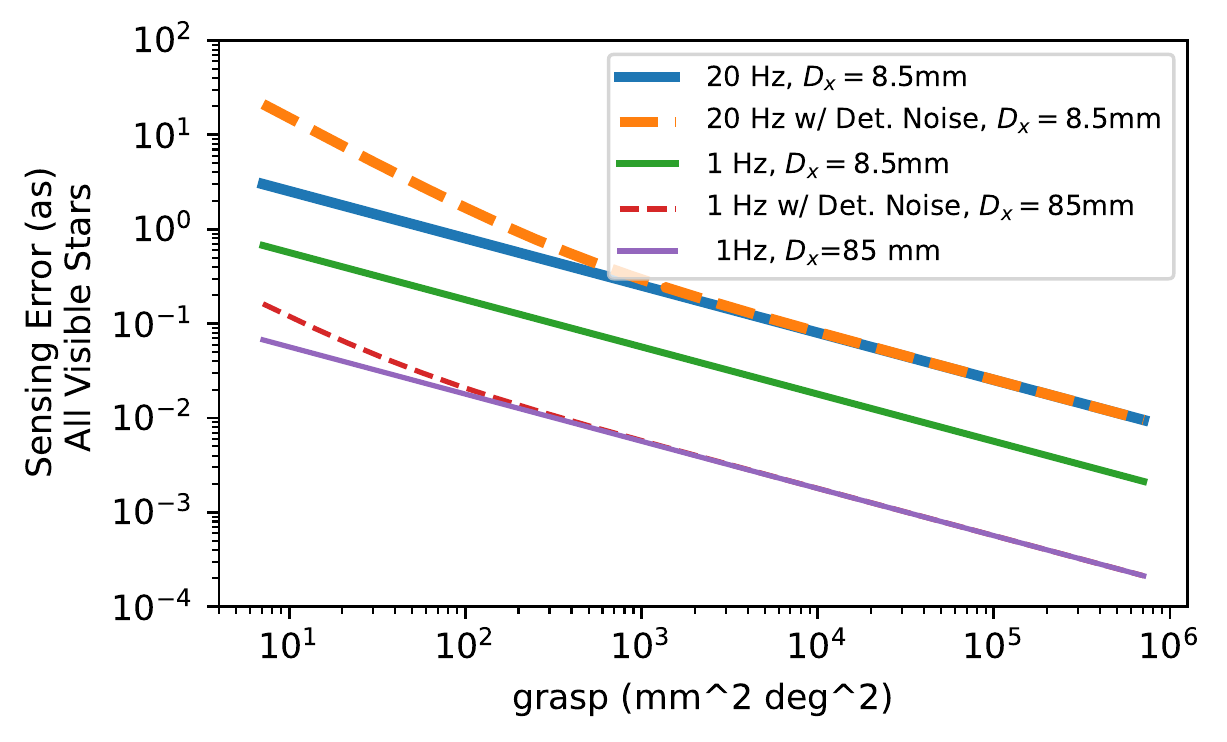}
    \caption{Attitude determination error versus system grasp or \textit{\'{e}tendue}.
    This photon noise limited case assumes the flux per square degree at the galactic poles for stars from zero to 13th magnitude, (Fig. \ref{fig:star_density}).
    The upper bound on the x-axis corresponds to a 85mm diameter telescope with a ten degree circular \gls{FOV}.
 {The} 20 Hz {curve is} shown to emphasize the importance of short exposure times.
 At low grasps the detector noise,  dominates while, at large grasps, the photon noise dominates.
 Small values of $D_x$ show the importance of sensor aperture to constraining body pointing, note the largest grasps are likely require {an} impractically large \gls{FOV} for  $D_x$=8.5mm.
    }
    \label{fig:sensing_vs_etendue}
\end{figure}

For our analysis, we assume the spacecraft is sufficiently stable that individual stars can be identified and smearing in a given exposure is negligible.
While not discussed here, the length of star smearing can be used to determine the angular rate of the spacecraft \citep{enright_towards_2010}.
Attitude sensing using astronomical sources requires sufficient photon counts to accurately determine direction by accurately measuring the star position(s). 
Fig. \ref{fig:sensing_vs_etendue} shows the sensing limits with (dashed curves) and without the addition of uncertainty due to a low-performance, high-\gls{TRL} detector (dot-dashed curves) described in Section \ref{sec:results}.

\begin{figure}[htp]
     \centering
    \includegraphics[width=0.48\textwidth]{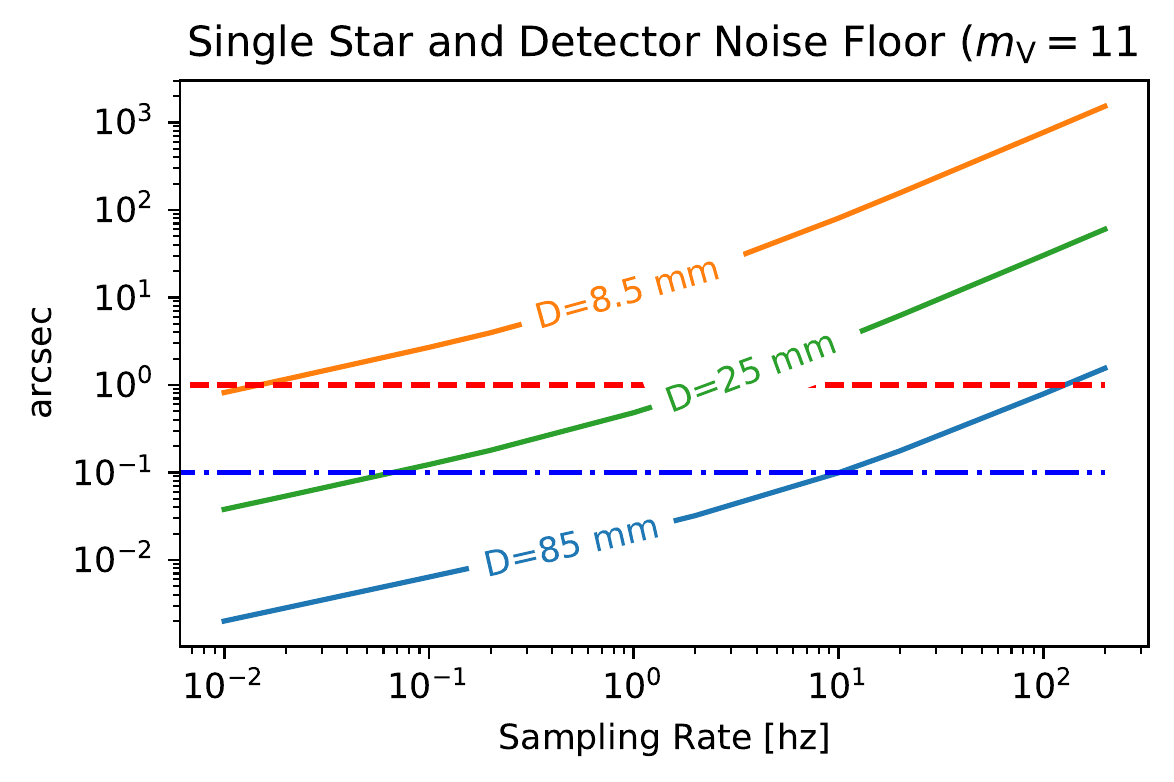}
    \caption{Attitude sensing limit curves assuming centroiding noise due to photon statistics and detector noise from the \gls{TRL}-9 MT9P031 CMOS detectors widely used on nanosatellites.  For an  85 mm space telescope, sub-arcsec sensing is readily achieved at 100 Hz sampling or slower rates, implying $\sim$10 Hz bandwidth controllers are as fast as is feasible on dim stars.}
    \label{fig:control_11_mag}
\end{figure}
\begin{figure}[htp]
\centering
    \setlength{\figureheight}{2.0in}
    \setlength{\figurewidth}{2.75in}
    \input{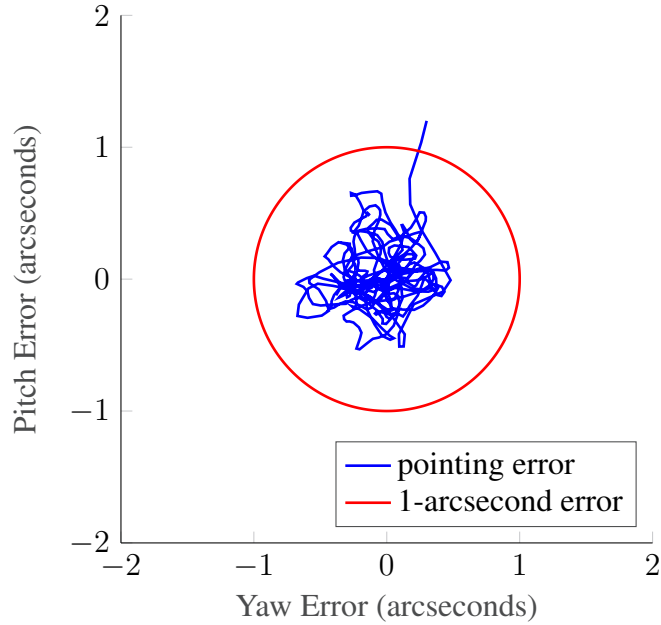}
    \caption{1 Hz closed-loop control simulation with disturbances from Fig. \ref{fig:disturbances} and sensor noise corresponding to the largest practical telescope that might fit in a 6U CubeSat with an 85 mm diameter, measuring the angle of a single $m_{\rm V}$= 11 star. The red circle indicates 1 arcsecond error. The initial state is outside the 1 arcsecond circle, but the controller quickly recovers. The resulting RMS body-pointing error is {0.39}\arcsec.}
    \label{fig:closed-loop}
\end{figure}

\section{State Estimation}\label{sec:estimation}

In addition to the astronomical telescope, a compact star tracker and gyroscope are included in our model for coarse attitude determination.
 Typical COTS star trackers for CubeSats provide attitude determination with an accuracy of tens of arcseconds{,} and are able to maintain tracking at slew rates of several degrees per second.
Measurements from all sensors are fused in a multiplicative extended Kalman filter (MEKF)~\citep{Lefferts82} to calculate a maximum-likelihood estimate of of the spacecraft's attitude.

In addition to estimating the attitude of the spacecraft body, parameterized by quaternion $q$, the filter also estimates a gyroscope bias vector $b${,} an external bias torque $\tau$\added{, and the time derivative of this external torque $\dot{\tau}$}. \added{We assume that the environmental disturbances vary slowly compared to the filter update rate, making a first-order process model sufficiently accurate.} Both {the gyro bias, as well as the derivative of the external torque,} are assumed to follow a random walk process, while the gyroscope measurements $\omega$ are assumed to be corrupted by additive white Gaussian noise. The MEKF process model is,
\begin{equation}
    \dot{x}_{kf} = \begin{bmatrix}
    \dot{q} \\ \dot{b} \\ \dot{\tau} \\ \added{\Ddot{\tau}}
    \end{bmatrix} = 
    \begin{bmatrix}
    \frac{1}{2} q \otimes (\omega + \nu_\omega) \\
    \nu_b \\ 
    \dot{\tau}\\ 
    \added{\nu_{\dot{\tau}}}
    \end{bmatrix} ,
\end{equation}
where $\nu_\omega$, $\nu_b$, and $\nu_{\dot{tau}}$ are noise inputs drawn from multivariate Gaussian distributions. The covariances $V_{\omega \omega}$ and $V_{b b}$ corresponding to the noise terms $\nu_\omega$ and $\nu_b$ are properties of the gyroscope used, while
$V_{\dot{\tau} \dot{\tau}}$
corresponding to $\nu_{\dot{\tau}}$ is calculated based on the expected environmental disturbance torques. The MEKF measurement model, which maps the state $x_{kf}$ into the expected sensor measurements $y$, is similarly given by,
\begin{equation}
    y = h(x_{kf}) + w ,
\end{equation}
where the function $h(x)$ returns expected star locations in the sensor frames based on the current spacecraft state and $w$ is assumed to be drawn from a multivariate Gaussian distribution with covariance matrix $W$ calculated based on Fig. \ref{fig:control_11_mag}.

\section{Results}\label{sec:results}
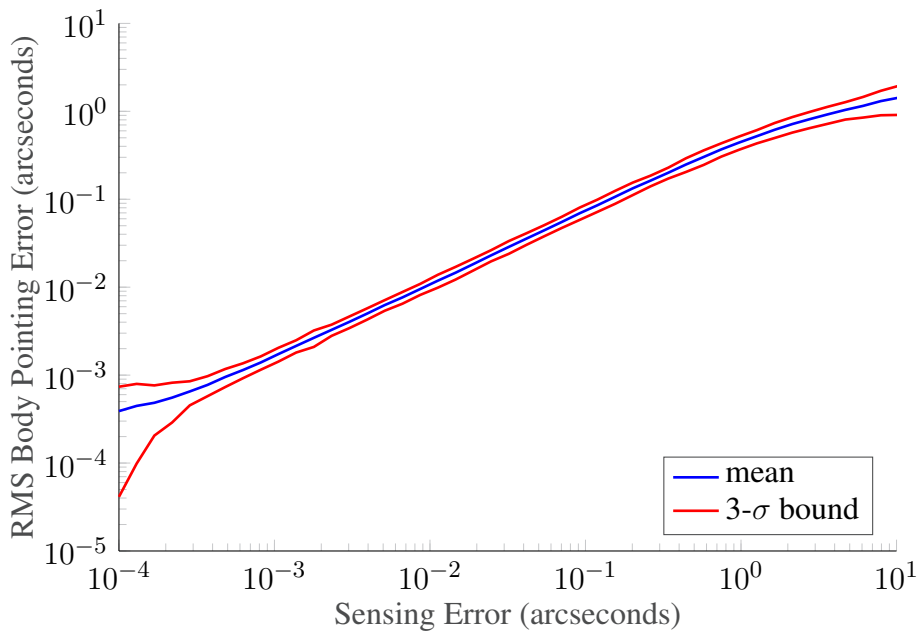
\begin{figure}[h!]
    \centering
%
%
\begin{tikzpicture}

\begin{axis}[%
width=4.028in,
height=2.754in,
at={(1.011in,0.642in)},
scale only axis,
xmode=log,
xmin=0.0001,
xmax=10,
xminorticks=true,
xlabel style={font=\color{white!15!black}},
xlabel={Sensing Error (arcseconds)},
ymode=log,
ymin=1e-05,
ymax=10,
yminorticks=true,
ylabel style={font=\color{white!15!black}},
ylabel={RMS Body Pointing Error (arcseconds)},
axis background/.style={fill=white},
axis x line*=bottom,
axis y line*=left,
legend style={at={(0.97,0.03)}, anchor=south east, legend cell align=left, align=left, draw=white!15!black}
]
\addplot [color=blue, line width=1.0pt]
  table[row sep=crcr]{%
0.0001	0.000389699268083956\\
0.00013	0.000446744232790638\\
0.000169	0.000484629520152547\\
0.0002197	0.000554504194847889\\
0.00028561	0.000652338047567351\\
0.000371293	0.000774079000015018\\
0.0004826809	0.000952992601182336\\
0.00062748517	0.00114226541572323\\
0.000815730721	0.00139108537920563\\
0.0010604499373	0.00173512582706282\\
0.00137858491849	0.00214818630458063\\
0.001792160394037	0.00265524904880938\\
0.0023298085122481	0.00327182228655399\\
0.00302875106592253	0.00403012618635312\\
0.00393737638569929	0.00500374315361651\\
0.00511858930140908	0.00626185655024364\\
0.0066541660918318	0.00765236199902353\\
0.00865041591938134	0.00952783457940181\\
0.0112455406951957	0.0118670588796381\\
0.0146192029037545	0.0146227555016306\\
0.0190049637748808	0.0183155100393309\\
0.0247064529073451	0.0230017307847768\\
0.0321183887795486	0.0287034803725272\\
0.0417539054134132	0.0356993485556443\\
0.0542800770374371	0.0443342745075533\\
0.0705641001486682	0.0552799392079227\\
0.0917333301932687	0.0693230015593861\\
0.119253329251249	0.0854068670459826\\
0.155029328026624	0.106634750931388\\
0.201538126434611	0.13321573750047\\
0.261999564364995	0.162905879019097\\
0.340599433674493	0.200265265343455\\
0.442779263776841	0.248430547366955\\
0.575613042909894	0.302768397623503\\
0.748296955782862	0.369498511468618\\
0.97278604251772	0.441488284649815\\
1.26462185527304	0.521533670335637\\
1.64400841185495	0.616502082212574\\
2.13721093541143	0.716433484768612\\
2.77837421603486	0.816219346773942\\
3.61188648084532	0.923718623432179\\
4.69545242509892	1.03948454854363\\
6.10408815262859	1.15390729822958\\
7.93531459841717	1.30755926362036\\
10.3159089779423	1.42910653272862\\
13.410681671325	1.60514275911257\\
};
\addlegendentry{mean}

\addplot [color=red, line width=1.0pt]
  table[row sep=crcr]{%
0.0001	4.1266922336251e-05\\
0.00013	9.87434079185999e-05\\
0.000169	0.000205448727879796\\
0.0002197	0.000289169521136456\\
0.00028561	0.000453417233532464\\
0.000371293	0.000575972349878715\\
0.0004826809	0.000732094516755243\\
0.00062748517	0.000921938288689614\\
0.000815730721	0.00115066001096124\\
0.0010604499373	0.00142517298620004\\
0.00137858491849	0.00180584861247615\\
0.001792160394037	0.00209306428732337\\
0.0023298085122481	0.00279555917894932\\
0.00302875106592253	0.00342139355335337\\
0.00393737638569929	0.00427027118671443\\
0.00511858930140908	0.00539819580709308\\
0.0066541660918318	0.0064889234039766\\
0.00865041591938134	0.00816187511651412\\
0.0112455406951957	0.00988180770518229\\
0.0146192029037545	0.0122166931867218\\
0.0190049637748808	0.0154919191727054\\
0.0247064529073451	0.0196619363726832\\
0.0321183887795486	0.0239520254520223\\
0.0417539054134132	0.0303215552302929\\
0.0542800770374371	0.0379453381365189\\
0.0705641001486682	0.047250597422561\\
0.0917333301932687	0.0582401608326919\\
0.119253329251249	0.0718423682277885\\
0.155029328026624	0.0887290650426445\\
0.201538126434611	0.111606102081419\\
0.261999564364995	0.140260820617985\\
0.340599433674493	0.171886010828405\\
0.442779263776841	0.204383342469712\\
0.575613042909894	0.245896412264154\\
0.748296955782862	0.305839459178871\\
0.97278604251772	0.365837486946624\\
1.26462185527304	0.431957799314264\\
1.64400841185495	0.497051130186377\\
2.13721093541143	0.572906261972058\\
2.77837421603486	0.645643037644474\\
3.61188648084532	0.721302632889339\\
4.69545242509892	0.806471887224974\\
6.10408815262859	0.84936730571409\\
7.93531459841717	0.902232408727491\\
10.3159089779423	0.910320446447343\\
13.410681671325	0.948498564810616\\
};
\addlegendentry{3-$\sigma$ bound}

\addplot [color=red,  line width=1.0pt,forget plot]
  table[row sep=crcr]{%
0.0001	0.000738131613831661\\
0.00013	0.000794745057662677\\
0.000169	0.000763810312425298\\
0.0002197	0.000819838868559322\\
0.00028561	0.000851258861602238\\
0.000371293	0.000972185650151321\\
0.0004826809	0.00117389068560943\\
0.00062748517	0.00136259254275684\\
0.000815730721	0.00163151074745001\\
0.0010604499373	0.0020450786679256\\
0.00137858491849	0.0024905239966851\\
0.001792160394037	0.00321743381029539\\
0.0023298085122481	0.00374808539415865\\
0.00302875106592253	0.00463885881935287\\
0.00393737638569929	0.00573721512051859\\
0.00511858930140908	0.0071255172933942\\
0.0066541660918318	0.00881580059407045\\
0.00865041591938134	0.0108937940422895\\
0.0112455406951957	0.013852310054094\\
0.0146192029037545	0.0170288178165395\\
0.0190049637748808	0.0211391009059565\\
0.0247064529073451	0.0263415251968704\\
0.0321183887795486	0.0334549352930322\\
0.0417539054134132	0.0410771418809957\\
0.0542800770374371	0.0507232108785876\\
0.0705641001486682	0.0633092809932843\\
0.0917333301932687	0.0804058422860803\\
0.119253329251249	0.0989713658641766\\
0.155029328026624	0.124540436820131\\
0.201538126434611	0.154825372919522\\
0.261999564364995	0.185550937420209\\
0.340599433674493	0.228644519858505\\
0.442779263776841	0.292477752264198\\
0.575613042909894	0.359640382982853\\
0.748296955782862	0.433157563758365\\
0.97278604251772	0.517139082353006\\
1.26462185527304	0.611109541357009\\
1.64400841185495	0.73595303423877\\
2.13721093541143	0.859960707565166\\
2.77837421603486	0.98679565590341\\
3.61188648084532	1.12613461397502\\
4.69545242509892	1.27249720986229\\
6.10408815262859	1.45844729074507\\
7.93531459841717	1.71288611851323\\
10.3159089779423	1.94789261900989\\
13.410681671325	2.26178695341452\\
};
\end{axis}
\end{tikzpicture}%
    \caption{Body pointing RMS error as it relates to sensing error (whether photon or detector limited).     Data are from 1000 Monte-Carlo trials with full environmental disturbance torques. Each point on these curves corresponds to a full simulation analogous to that shown in Fig. \ref{fig:closed-loop}.}
    \label{fig:pointing_mc}
\end{figure}

 Fig. \ref{fig:control_11_mag} shows, for example, a single $m_{\rm V}$= 11 guide star can provide sub-0.1\arcsec pointing knowledge to a 1 Hz sampling system, assuming typical detector noise levels in Table \ref{tab:nomenclature} to the calculation of  $\sigma$.
 The noise characteristics assumed approximate the MT9P031 \gls{CMOS} sensor with 2.2$\mu$m pixels, commonly used for precision nanosatellite applications \citep{becker_commercial_2008,enright_things_2012-1,allan2018deformable}.
$m_{\rm V}$ = 11 was chosen because they are common, $>>1/$star sq. deg on average.
Due to the extremely low frequency content of on-orbit environmental disturbances (Fig. \ref{fig:disturbances}), the example sampling at 1 Hz is a far higher sample rate than would be needed for an idealized control loop.

Spacecraft body pointing, applying the control system (Fig. \ref{fig:control_block_diagram}) to input the disturbances, including photon and sensor noise for a 85 mm telescope observing a single $m_{\rm V}$ star sampled at 1 Hz is simulated in Fig. \ref{fig:closed-loop}.
This figure shows the time evolution of body pointing for the proposed control system{, with a one arcsecond circle for reference}. With the starting point outside the one-arcsecond circle, the  body pointing of the reference 6U CubeSat orbit is quickly controlled to an RMS of {0.39} arcseconds.

Fig. \ref{fig:pointing_mc} extends this analysis to a range of sensing errors. 
The center line, bracketed by $3-\sigma$ bounds, shows the RMS body pointing versus absolute sensing error, \added{demonstrating a relationship where the pointing error closely matches the sensing error over a broad range. 
This is noteworthy since it suggests that the limiting factor in pointing performance comes from the sensing errors instead of jitter induced by the actuators or bandwidth limits in the control system.}
The sensor grasp or aperture that would reach this level of performance can be found by observing the sensing errors in Fig. \ref{fig:control_11_mag} or \ref{fig:sensing_vs_etendue}. 
For example, to reach 0.1\arcsec at 1 Hz using all available starlight, a grasp of $\gtrsim$200  mm $\deg^2$ is needed, while a single 11th magnitude star is insufficient to reach that level in a nanosatellite aperture.

\section{Conclusions}\label{sec:conclusions}

In 1980 Nancy Grace Roman said  ``pointing has been the pacing team that has really controlled what we've been able to do in space astronomy as the field has developed''\citep{roman_oral_1980}\footnote{\url{www.aip.org/history-programs/niels-bohr-library/oral-histories/4846}}.
This assertion, a ``Roman's Law" for spacecraft capabilities, continues to hold true, and pointing remains a particular challenge for nanosatellites. 

Detector dynamic range constraints were neglected in this analysis: Present-day sensors with limited bit depth will not be able to capture all the incident photons without saturating. This limits the useful input flux. 
Similarly, low noise detectors allow sensing of dimmer stars, changing the brightness cutoff, $m_l$.
Since the brightest stars are also the rarest, this will generally have less impact than might be first assumed{,} but suggests the importance of both improved actuators and high-dynamic-range sensors and readout electronics for future \glspl{ADCS}.

Detector pixel sampling was implicitly optimized across this analysis and, for physical designs, the \gls{FOV} and grasp must be carefully weighed against pixel size and detector noise levels. 
Caution must be used {when} applying the sensing-error-versus-grasp curve on Fig. \ref{fig:sensing_vs_etendue} to constrain physical designs. 
Physical designs must include detector noise{,} pixel sampling values{,} and centroiding precision.
With sufficient flux rates, increasing detector noise by defocusing stars improves centroiding to millipixel levels \citep{buffington_using_1991}{,} but must be balanced with per-pixel noise contributions and confusion limits. 

The closed-loop simulation results presented here show that the low frequency of environmental disturbances allows for a very slow control loop {for} inertially pointing spacecraft.
While further improvements in estimation and control algorithms are possible, this work demonstrates {practical} limits of nanosatellite pointing far beyond the current state-of-the-art{,} and establishes the target for actuator improvements to enable precision astrophysics with nanosatellite{s}.

\section*{Conflict of Interest Statement}

The authors declare that the research was conducted in the absence of any commercial or financial relationships that could be construed as a potential conflict of interest.

\section*{Author Contributions}
E.S.D. led experiment design, simulation, and manuscript preparation.
Z.M. led model design, optimization,  analysis, and contributed  to manuscript preparation.
K.T. led disturbance simulations, and contributed to literature review and manuscript preparation.

\section*{Funding}
Portions of this work were supported by the Arizona Board of Regents Technology Research Initiative Fund (TRIF).

\section*{Acknowledgments}
The authors thank Laurent Pueyo for many helpful discussions and feedback.
This research has made use of the the VizieR catalogue access tool, CDS, Strasbourg, France (DOI : 10.26093/cds/vizier). The original description 
 of the VizieR service was published in 2000, A\&AS 143, 23.

\section*{Data Availability Statement}
Code to reproduce the figures in this paper is available on \url{https://github.com/RoboticExplorationLab/FinePointing} and citable and archived by Zenodo \citep{ewan_douglas_douglasefinepointing_2021}.

\bibliographystyle{frontiersinSCNS_ENG_HUMS} 
\bibliography{test}

\appendix

\end{document}